\documentclass[aps,prb,reprint,amsmath,amssymb,floatfix,superscriptaddress]{revtex4-1}

\usepackage{graphicx}
\usepackage{soul} 
\usepackage{xcolor} 

\begin{document}
\title{Thermal conductance of thin film YIG determined using Bayesian statistics}
\author{C. Euler}
\email{eulerch@uni-mainz.de.}
\affiliation{Institut f\"ur Physik, Johannes Gutenberg-Universit\"at Mainz, Staudinger Weg 7, 55128 Mainz, Germany}
\author{P. Ho\l{}uj}
\affiliation{Institut f\"ur Physik, Johannes Gutenberg-Universit\"at Mainz, Staudinger Weg 7, 55128 Mainz, Germany}
\affiliation{Graduate School of Excellence \lq Materials Science in Mainz\rq , Staudinger Weg 9, 55128 Mainz, Germany}
\author{T. Langner}
\affiliation{Fachbereich Physik und Landesforschungszentrum OPTIMAS, Technische Universit\"at Kaiserslautern, Erwin-Schr\"odinger-Stra\ss e 56 , 67663 Kaiserslautern, Germany}
\author{A. Kehlberger}
\affiliation{Institut f\"ur Physik, Johannes Gutenberg-Universit\"at Mainz, Staudinger Weg 7, 55128 Mainz, Germany}
\affiliation{Graduate School of Excellence \lq Materials Science in Mainz\rq , Staudinger Weg 9, 55128 Mainz, Germany}
\author{V. I. Vasyuchka}
\affiliation{Fachbereich Physik und Landesforschungszentrum OPTIMAS, Technische Universit\"at Kaiserslautern, Erwin-Schr\"odinger-Stra\ss e 56 , 67663 Kaiserslautern, Germany}
\author{M. Kl\"{a}ui}
\affiliation{Institut f\"ur Physik, Johannes Gutenberg-Universit\"at Mainz, Staudinger Weg 7, 55128 Mainz, Germany}
\affiliation{Graduate School of Excellence \lq Materials Science in Mainz\rq , Staudinger Weg 9, 55128 Mainz, Germany}
\author{G. Jakob}
\affiliation{Institut f\"ur Physik, Johannes Gutenberg-Universit\"at Mainz, Staudinger Weg 7, 55128 Mainz, Germany}
\affiliation{Graduate School of Excellence \lq Materials Science in Mainz\rq , Staudinger Weg 9, 55128 Mainz, Germany}

\date{\today}

\begin{abstract}
Thin film YIG (Y$_3$Fe$_5$O$_{12}$) is a prototypical material for experiments on thermally generated pure spin currents and the spin Seebeck effect. The 3$\omega$ method is an established technique to measure the cross-plane thermal conductance of thin films, but can not be used in YIG/GGG (Ga$_3$Gd$_5$O$_{12}$) systems in its standard form. We use two-dimensional modeling of heat transport and introduce a technique based on Bayesian statistics to evaluate measurement data taken from the 3$\omega$ method. Our analysis method allows us to study materials systems that have not been accessible with the conventionally used 3$\omega$ analysis. Temperature dependent thermal conductance data of thin film YIG are of major importance for experiments in the field of spin-caloritronics. Here we show data between room temperature and 10~K for films covering a wide thickness range as well as the magnetic field effect on the thermal conductance between 10~K and 50~K.
\end{abstract}

\maketitle

Knowledge of the thermal properties of a system is fundamental in experiments including heat flow. In solids, heat is mainly transported by long-wavelength acoustic phonons. Therefore, even comparably thick 'thin film' systems encounter the generic issue that the heat conductivity depends on the film thickness, since long wavelength phonons will be scattered by the surfaces or interfaces. For the electrical conductance such an effect is described by the Fuchs-Sondheimer relation.\cite{Fuchs1938, Sondheimer2001} The derivation of this relation was simplified by the comparably easy access to experimental data, i.e.\ measurement of electrically conducting thin films on insulating substrates. In addition, due to the Pauli principle only electrons at the Fermi surface participate in transport, so that a Fermi velocity and mean free path for the theoretical analysis can be computed. For heat transport the full phonon spectrum needs to be considered and the heat conduction of the substrate can usually not be eliminated in the experiment rendering the determination of thin film heat conductance an ambitious task.

\par The interaction of the spin degree of freedom with an energy flow in a solid, spin-caloritronics, is a recent field of research\cite{Bauer2012,Boona2014} in which YIG (Y$_3$Fe$_5$O$_{12}$) is of interest due to its low magnetic damping. Temperature driven pure spin currents and the spin Seebeck effect have been observed using YIG based samples\cite{Uchida2010,Schreier2013,Kehlberger2013} and there is an ongoing discussion about the origin of the spin Seebeck effect,\cite{Jaworski2011,Gepraegs2014} which includes magnon-phonon interaction. Here the so far unknown temperature dependence of the thermal conductance of thin film YIG is needed to quantitatively determine the real temperature gradients in the investigated systems. While values of the bulk thermal conductivity vary between $(6\dots8)\mathrm{~W/(m\,K)}$\cite{Slack1971,Padture1997,Hofmeister2006}, Padture et al.\cite{Padture1997} discuss that their result of approx.\ 6~W/(mK) possibly underestimates the true value by up to 1.5~W/(mK). Theoretical calculations of the magnon thermal conductivity in bulk YIG exist,\cite{Rezende2014} but vastly disagree in magnitude with experimental results.\cite{Boona2014b}

\par Mostly for numerical reasons conventional methods to determine thin film thermal conductances fail in systems in which the film thermal conductivity is comparable to that of the substrate. Therefore, crucial temperature gradients are usually calculated using bulk thermal conductivities\cite{Schreier2013,Boona2014b} even if films of several hundred nm thickness are used. In this paper we introduce a data evaluation scheme based on Bayesian statistics to determine the thickness dependent heat conductance of thin film samples. Our novel method allows us to determine the thermal conductivity of systems not only in which the substrate has a heat conductivity much larger than that of the film material, but independent of this condition. We present the thickness dependent thermal conductivity of thin film YIG and demonstrate that at temperatures far below room temperature the thermal conductance is mainly limited by point defects.

\par
\begin{figure}
	\centering
	\includegraphics[width=\linewidth]{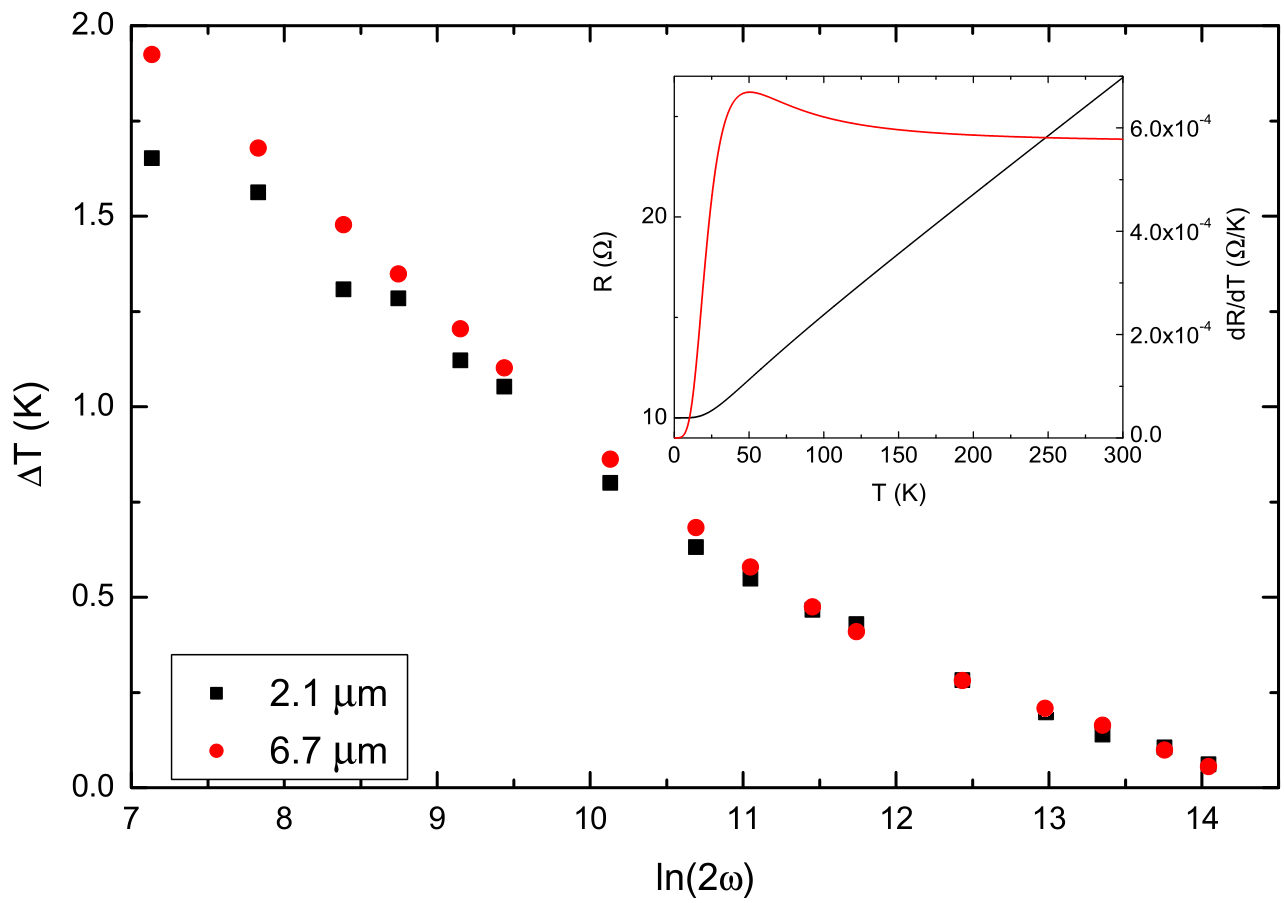}
	\caption{\label{fig:thermal_penetration_depth}The temperature oscillation in the heater structure in dependence of $\ln{(2\omega)}$ measured between 100~Hz to 100~kHz with $\omega$ the dimensionless angular frequency measured in Hz. The inset shows the temperature dependence of the resistance of the Au heater strip (black) and its temperature derivative (red).}
\end{figure}

 To determine the thermal conductance we applied the 3$\omega$ method that uses a patterned metallic structure as a heater and measurement device for the thermal resistance of the sample.\cite{Cahill1987,Cahill1990} An alternating current with frequency $\omega$ is applied to the heater structure, heating the sample and modulating the resistance $R$ with a frequency of 2$\omega$. Ohm's law and trigonometric identities imply a 3$\omega$ contribution to the voltage, $U_{3\omega}$, related to the temperature oscillation in the heater. $U_{3\omega}$ furthermore depends on the heater's temperature coefficient of resistance and the electric heating power applied. Generally, the thermal penetration (probing) depth of the experiment depends on the thermal diffusivity of the material and the applied frequency as $\omega^{-0.5}$, so that with increasing frequency the thickness of the layer \lq seen\rq\, decreases. While at low frequencies the experiment is sensitive to both film and substrate, at high frequencies the thermal penetration depth decreases until only the film is measured. The thermal conductivity can then be inferred from the slope of $U_{3\omega}$ with respect to $\ln{(2\omega)}$ (slope method).\cite{Cahill1987} Fig. \ref{fig:thermal_penetration_depth} shows the temperature oscillation in dependence of $\ln{(2\omega)}$. In the high-frequency range above 50~kHz the data of the $6.7~\mu$m and $2.1~\mu$m films overlap, indicating that the thermal penetration depth is smaller than the thickness of the thinner film (at 50,525~Hz the thermal penetration depth amounts to 2.1~$\mu$m). Particularly in this regime the finite heater width needs to be taken into account \cite{Cahill1990}. In the standard procedure at low frequencies the heat resistance of the film is compared to that of a reference sample with a smaller film thickness (differential method). This approach requires the thermal transport to be strictly perpendicular to the sample surface by assuming that the thermal conductance of the substrate by far exceeds that of the film and that the thermal penetration depth is larger than the film thickness. The first requirement concerns the choice of the substrate. The second can only be met by an adequate heater structure and measurement frequency.

\par For YIG the choice of substrates is largely limited to other garnets. However, the traditional 3$\omega$ method is not usable for YIG films on Gd$_{3}$Ga$_{5}$O$_{12}$ (GGG) substrates, as the thermal conductivities of film and substrate are comparable.\cite{Schreier2013} To conserve the validity of the classical $3\omega$ approach, the information depth of the method needs to be smaller than the film thickness, which requires  electron beam lithography and high-frequency detection (see Fig.~\ref{fig:thermal_penetration_depth}). In addition, for very high frequencies and very low temperatures the amplitude of the temperature oscillations gets very small and the slope based method becomes inapplicable (below approx. 50~K in the case of YIG). Thus, at sufficiently low temperatures not only do the assumptions of the differential 3$\omega$ method, e.g.\ the substrate as a heat sink, break down, but also the applicability of the slope method. This underlines the necessity for a new approach to determine thermal conductances.

The film's thermal conductance and the temperature oscillation in the heater are connected by the solution of the heat diffusion equation. Instead of remaining in the strict cross-plane (one-dimensional) transport regime, the requirements for the heater geometry and the substrate thermal conductivity can be relaxed by extending the equation for the temperature increase in the heater to a model of two-dimensional (in- and cross-plane) heat transport.\cite{Carslaw1959,Feldman1996,Borca-Tasciuc2001} For an arbitrary system of films on a substrate, the solution for the complex temperature oscillation in a heater line is given by the integral equation\cite{Borca-Tasciuc2001}

\begin{eqnarray}
	\Delta T = -\frac{P}{\pi l \kappa_{z,1}}\int^\infty_0 dk\, \frac{1}{A_1 B_1}\frac{\sin^2{(bk)}}{(bk)^2}\label{eq:DT}\\
	B_i = \sqrt{\frac{\kappa_{xy,i}}{\kappa_{z,i}}k^2 + \frac{2i\omega}{\alpha_{z,i}}}\label{eq:DT_Bi}\\
	A_{i-1} = \frac{A_i\frac{\kappa_{z,i}B_i}{\kappa_{z,i-1}B_{i-1}}-\tanh{(B_{i-1}d_{i-1})}}{1-A_i\frac{\kappa_{z,i}B_i}{\kappa_{z,i-1}B_{i-1}}\tanh{(B_{i-1}d_{i-1})}}\\
	A_n = -1.
\end{eqnarray}

\noindent where $P$ is the applied heating power and the heater stripe has the dimensions $2b$ and $l$. $A_1$ and $B_1$ are given by recursive sequences $A_i$ and $B_i$ depending on the in-plane and cross-plane thermal conductivities $\kappa_{xy,i}$ and $\kappa_{z,i}$, the thickness $d_i$ of the \textit{i}-th layer of the stack and the respective thermal diffusivities $\alpha_i=\frac{\kappa_{z,i}}{\rho_i c_{s,i}}$ given by the thermal conductance, the density $\rho$ and the specific heat $c_s$. As arbitrary thicknesses $d_i$ can enter the calculation, this method is independent of the thickness of the individual layers. However, as the thermal \lq wave\rq\, decays exponentially within a layer, layers thicker than approx. 100~$\mu$m are difficult to analyze. Interface thermal (Kapitza) resistances are neglected here, but can in principle be introduced.\cite{Carslaw1959,Olson2005,Kimling2013thesis}

\par The real part of the temperature oscillation resembles an improved version of the traditional approach by including the anisotropy of the thermal conductance. Therefore, in principle it is possible to determine also the in-plane contribution to thermal transport without a second 3$\omega$ structure and without varying the width of the heater line. Furthermore, the real part depends on the density and the specific heat only by cross-terms introduced by the multiplication of the complex quantities $A_i$ and $B_i$. The imaginary part in particular includes an $\alpha_{z,i}^{-1}$ term and therefore is a measure for the density and the specific heat, but is also sensitive to the cross-plane thermal conductance. 

\par If for each layer $\{\kappa_{xy}, \kappa_z, \rho, c_s\}$ are unknown, the equation contains $4\cdot N_\mathrm{layers}$ free parameters and it could theoretically possess a unique solution, if the measured set of frequencies was equal to or greater than the number of parameters. The large number of parameters then leads to a high degree of computational complexity and instability. Assuming all properties of all layers except for that of interest are known from previous measurements, the number of parameters can be reduced to those of the film. A sensitivity analysis of Eq.~\ref{eq:DT} to the material parameters allows for further simplification. We found that any determination of $\rho$ and $c_s$ will be prone to large uncertainties and is only possible for materials with very large density and specific heat capacity. Thus, determining the thermal conductivities of the film can be simplified by assuming reasonable values for the film's density and temperature dependent specific heat capacity\cite{Hofmeister2006} and only determining $\kappa_{xy}$ and $\kappa_z$.

\par Fitting Eq.~\ref{eq:DT} to measurement data is complex due to the  integral equation. Therefore, techniques such as a differential material properties search algorithm\cite{Olson2005,Kimling2013thesis} or neural networks\cite{Feuchter2014} are implemented to find a meaningful fit. It may also be possible to use the Monte Carlo technique to obtain a solution, which, however, does not need to be unique. E.g.\ the algorithm presented by Olson et al.\cite{Olson2005} lacks stability and the results depend on the initial conditions of the fit. To circumvent this difficulty, we have implemented a new algorithm based on Bayes' theorem, Eq.~\ref{eq:bayes}. Bayesian statistics uses previously determined information (\lq prior probability distribution\rq) and then uses data to compute a \lq posterior distribution\rq .

\begin{equation}
	p(m_i|o) = \frac{p(o|m_i)p(m_i)}{\sum_j p(o|m_j)p(m_j)}
	\label{eq:bayes}
\end{equation}

\noindent Here $m_i$ denotes the \textit{i}-th model of a series of models, $o$ a measurement and $p(m_i|o)$ the conditional probability that a model $m_i$ describes the physical reality provided the observation $o$. Vice versa, $p(o|m_i)$ is the non-normalized likelihood that an observation $o$ be made provided $m_i$ is real. The denominator is the effective normalization of the probability $p(m_i|o)$ to $[0,1]$. Assuming an experiment is carried out and the outcome is compared to $N_m$ models of the process, as a prior one could choose a uniform distribution. The difference between model value $m_i$ and observational value $o$ can be expressed in terms of the uncertainty of the observation $\sigma$ by $\Sigma_i = \left\vert\frac{m_i-o}{\sigma}\right\vert$. This number represents an approximation for the probability that the observed value actually does not come from that model, so that

\begin{equation}
	p(o|m_i) = 2-2\phi(\Sigma_i),
	\label{eq:erf}
\end{equation}

\noindent where $\phi(\Sigma)$ is the cumulative Gaussian distribution running from 0.5 to 1, as the $\Sigma_i$ are positively defined.

\par For each combination of the parameters $\{\kappa_{xy}, \kappa_z, \omega\}$ within a physically meaningful range the expected $\Delta T$ is calculated to form an extensive catalog of models for temperatures between room temperature and 10~K. As a naive prior a uniform probability distribution function (PDF) for $\kappa_{xy}$ and $\kappa_z$ is chosen and then improved by disfavoring a large anisotropy in the thermal conductance.\cite{Boona2014b} Each experimental data point at a specific frequency is compared to all models at that frequency using Eq.~\ref{eq:bayes} and Eq.~\ref{eq:erf}. This leads to an \lq intermediate\rq\, PDF, which is recursively used as a prior for the next frequency data set and thereby improves the total PDF with increasing amount of data. The PDF determined after all data has been included represents the posterior PDF for $\kappa_{xy}$ and $\kappa_z$ from which the expectation value and the variance are computed.

\par To determine the credibility of our analysis method, we re-evaluated data taken on La$_{0.67}$Ca$_{0.33}$MnO$_3$\cite{Euler2015}, which had been evaluated using the conventional differential 3$\omega$ method. Our results agree very well within the experimental uncertainty and reproduce the results obtained in the cited reference. In addition, we compared a sample consisting of Au/Al$_2$O$_3$/MgO to a second sample Au/MgO to determine possible interface contributions. The results reproduce literature values of the thermal conductivity of Al$_2$O$_3$ without the need for an additional thermal interface resistance and therefore Kapitza resistances can be neglected.

\begin{figure}
	\centering
	\includegraphics[width=\linewidth]{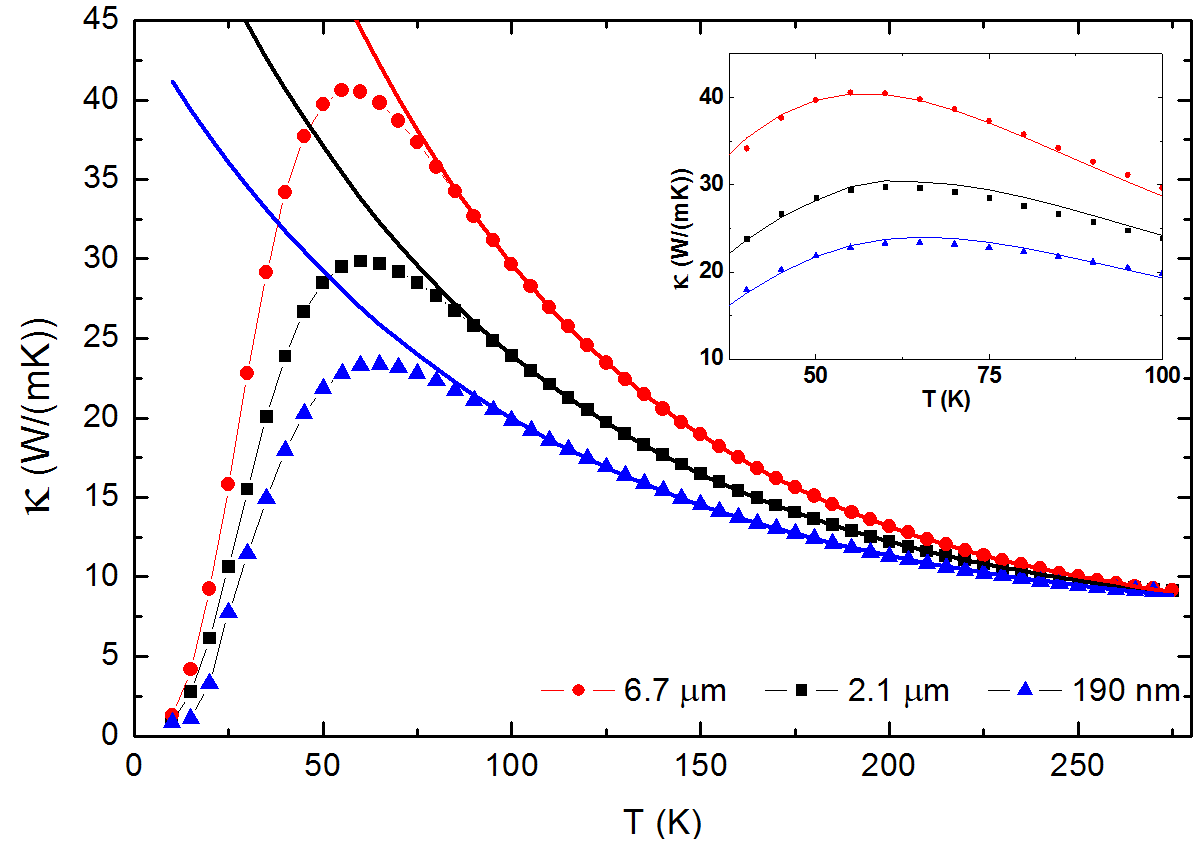}
	\caption{\label{fig:k_Tdep}Temperature dependent thermal conductances of YIG samples (first batch: 6.7~$\mu$m (red circles), 2.1~$\mu$m (black squares) and 190~nm (blue triangles)) and exponential fits to the high-temperature regions (solid lines). The inset displays the fit by the Callaway model around the maximum.}
\end{figure}

\par We employed our analysis method to YIG grown on (5$\times$10)~mm GGG substrates. A first batch of samples was deposited by liquid phase epitaxy (LPE) and pulsed laser deposition (PLD). The film thicknesses were $6.7~\mu\mathrm{m}$ and $2.1~\mu\mathrm{m}$ (LPE) and $190~\mathrm{nm}$ (PLD). The second batch included samples grown by LPE with thicknesses of 1.51~$\mathrm{\mu m}$, 3.08~$\mathrm{\mu m}$, 7.78~$\mathrm{\mu m}$, 12.01~$\mathrm{\mu m}$, 22.83~$\mathrm{\mu m}$ and 50.66~$\mathrm{\mu m}$. The gold layer from which the 3$\omega$ heater structure was fabricated was grown by DC sputter deposition. AFM measurements show the YIG film roughness to be below 1~nm. X-ray diffraction spectra demonstrate that sputtered Au has a polycrystalline fcc structure and X-ray reflectivity results in a RMS roughness of 4~nm. The heater strip (thickness: 50~nm) is rectangular in shape with a length of 1~mm between the voltage contacts and a width of 20~$\mu$m, which yields a room temperature specific resistivity of Au of $2.6\cdot 10^{-8}~\Omega\mathrm{m}$. The temperature dependence of its resistance and temperature coefficient of resistance are shown in the inset of Fig.~\ref{fig:thermal_penetration_depth} and follow the Bloch-Gr\"{u}neisen law. In particular, the slope of the resistance with the temperature is a measure of of the signal resolution, so that below 10~K no meaningful 3$\omega$ measurement can be performed. Caused by the low temperature coefficient below 10~K a measurement of the thermal conductivity is only possible above this temperature. An \textit{Anfatec eLockIn 205/2} lock-in amplifier, assisted by a Wheatstone bridge and common-mode subtraction\cite{Kimling2011}, was used to apply the heating voltage and read out $U_{3\omega}$.

\par A slope-based measurement of the thermal conductance of a GGG substrate was used to supply the algorithm with the thermal conductance of the substrate. While the slope method yielded a cross-plane thermal conductance for GGG of $\kappa_z = 7.8$~W/(mK), the two-dimensional method resulted in $\kappa_z = (8.6\pm 1.6)$~W/(mK). The room temperature thermal conductance of YIG was determined for two samples from the high-frequency data, in which the reduced thermal penetration depth causes the method to be sensitive to the film only. The cross-plane thermal conductance of the films is thereby determined to $\kappa_z = (9.0\pm 0.2)$~W/(mK) at room temperature. The Bayesian fitting algorithm was then applied to the low-frequency data below 2~kHz and led to values of $\kappa_{xy} = (9.5 \pm 1.1) \mathrm{~W/(mK)}$ and $\kappa_z = (8.5 \pm 0.6) \mathrm{~W/(mK)}$, which demonstrates the reliability of our algorithm.

\par The Bayesian approach is capable of providing the temperature dependent thermal conductance down to low temperatures. Fig.~\ref{fig:k_Tdep} shows the thermal conductance of the first batch of YIG films and displays clear peaks. At high temperatures, Umklapp scattering is the dominant factor that reduces the thermal conductance and the thermal conductances are expected to be thickness independent, while the peak is a measure for the length scale at which phonon boundary scattering becomes predominant. Therefore, a smaller film thickness should lead to a shift of the peak towards higher temperatures and, correspondingly, to lower peak values of the thermal conductance, which is observed in the first batch of samples. In bulk material Boona et al.\cite{Boona2014b} found a peak thermal conductivity of 110~W/(mK) at 25~K. At high temperatures the data fit an exponential decay close to the maximum as is expected from phonon transport theory.

\begin{figure}
	\centering
	\includegraphics[width=\linewidth]{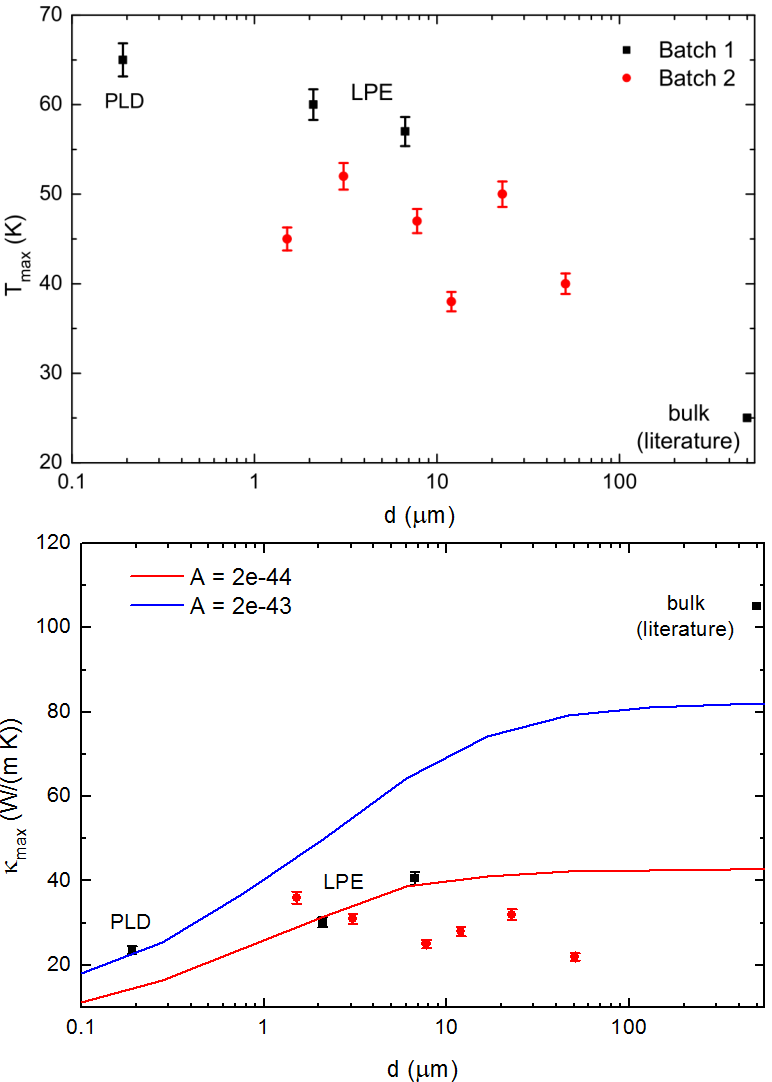}
	\caption{\label{fig:Tmax_kmax}Peak temperature $T_\mathrm{max}$ (upper plot) of the thermal conductance and peak value $\kappa_\mathrm{max}$ (lower plot) as function of film thickness. The data point at 500~$\mu$m is taken from Boona et al.\cite{Boona2014b} The black squares result from the first batch of samples, while the red circles correspond the second batch of samples. The error bars are derived from the sample-to-sample variation of $T_\mathrm{max}$ and $\kappa_\mathrm{max}$. The blue and red lines in the lower plot indicate the dependence on the defect concentration parameter of the maximum in the thermal conductivity at constant scattering parameters other than the thickness effect.}
\end{figure}

\par Modeling of the low-temperature behavior requires an in depth analysis of contributing transport processes. While we could experimentally determine the temperature dependence of the thermal conductivity, a detailed analysis of the different scattering contributions will require an independent and extensive determination of the defect structures for each sample, which is outside the scope of the current paper. Nevertheless, we have performed a phenomenological deconvolution of different scattering contributions in order to point out the challenges and possibilities. Specifically, thermal conductivity data can be modeled by contributions from different relaxation times $\tau$ as described by Callaway\cite{Callaway1959}. In particular, scattering of phonons at frequency $\omega$ at point defects is described by $\tau_d^{-1}=A\omega^4$ with $A$ a constant proportional to the defect density. The samples of the first batch grown by LPE can be explained using a single set of relaxation times and only varying the thickness-dependent boundary scattering relaxation time based on the actual thickness of the films (lower panel of Fig.~\ref{fig:Tmax_kmax}). Data taken from the film grown by PLD deviates from this model. A different parameter set is required in which the largest difference between the samples is a different defect concentration given by the model parameter $A$. The order of magnitude difference in the values of $A$ implies a large difference between the defect densities. The defects do not need to be of the same type in both sets of samples and a difference in film quality (PLD resulting in a larger defect concentration) is to be expected. In the case of the bulk sample data taken from literature there may be additional influences that are unknown to us.

\par It can be expected that films approaching the bulk state approach bulk values of $\kappa$, but this effect is unobserved in our data (Fig.~\ref{fig:Tmax_kmax}). There is no apparent systematic trend, even though all thermal conductivity curves can be fitted individually using the Callaway model, but with strongly varying defect concentrations. The sample-to-sample variation determined from two randomly chosen samples within one wafer of the maximum thermal conductivity amounts to 8\% and that of the maximum temperature to 6\%. This confirms the reproducibility of our results, as the 3$\omega$ method generally is prone to systematic uncertainties of up to 10\%. Our data imply that the quality of the film, i.e.\ the defect concentration, is a vital piece of information required to understand the thin film thermal conductance of YIG. At room temperature the values of the thermal conductance obtained using our model are slightly larger than the literature value. The fact that the slope-based values of the thermal conductivity of YIG and the result of the presented algorithm agree well at room temperature implies that our new method is capable of providing physically sound results.

\begin{figure}
	\centering
	\includegraphics[width=\linewidth]{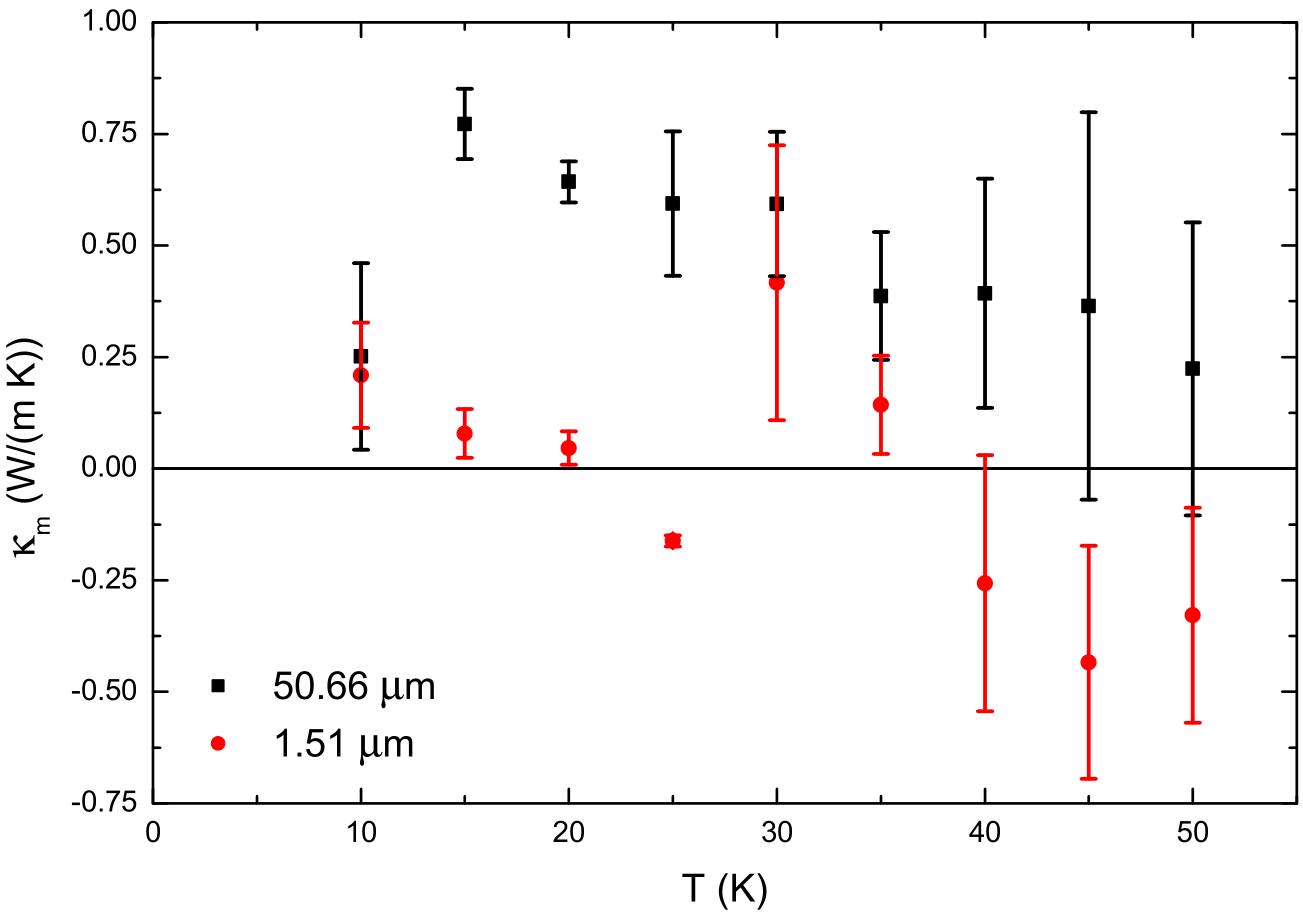}
	\caption{\label{fig:kmag_Tdep}Magnetic field dependent contribution of the thermal conductivity of YIG films with thicknesses of 50.66~$\mu$m and 1.51~$\mu$m.}
\end{figure}

\par Finally, Fig.~\ref{fig:kmag_Tdep} displays the difference in thermal conductivity $\kappa_m$ of YIG films between 10~K and 50~K in samples with thicknesses of 50.66~$\mu$m and 1.51~$\mu$m when in a magnetic field of 8~T, which freezes out the magnonic contribution,\cite{Boona2014b} and in zero field. Statistical errors are determined by sampling 100 data points per temperature point, which is necessary due to the small magnitude of the difference. Our data show that $\kappa_m$ may be thickness dependent and that the magnetic field may affect the thermal conductivity differently in thick films, while the thin film does not show a contribution significantly different from zero. Above 40~K the difference cannot be distinguished from thermal noise. The data suggest that the temperature of the largest difference is at approx. 15~K in the thick film, while there is no significant difference in the thin film at all. The point of largest difference is at a temperature much lower than the peak thermal conductivity, which is in agreement with \citet{Boona2014b}. Generally, we infer that the magnetic field dependent contribution to the thermal conductivity below 50~K does not exceed 1~W/(mK). Further research will be required to eliminate systematic uncertainties, which led to small but unphysical negative values of the thermal conductivity, and to understand the magnetic field effect on the thermal conductivity in YIG.

\par In conclusion we developed a method determining the thermal conductance of thin film samples based on two-dimensional heat transport and Bayesian statistics. We applied the method to YIG films on GGG substrates where the original $3\omega$ method is inapplicable. For thin film YIG we determined a room temperature thermal conductance of $\kappa_z = (8.5 \pm 0.6)~\mathrm{~W/(mK)}$ with low anisotropy. We present temperature dependent thermal conductances of thin film YIG between room temperature and 10~K with sample thicknesses between 190~nm and 50~$\mu$m. The results agree with bulk values at room temperature, but differ significantly towards lower temperature, where the thermal conductance is shown to be significantly reduced at small thicknesses compared to bulk material. In particular, in films grown by LPE the effect of the defect density exceeds that of the expected thickness dependence.

\par We acknowledge financial support by DFG SPP 1538 \lq Spin Caloric Transport\rq. P.H. and A.K. thank the Graduate School of Excellence \lq Materials Science in Mainz\rq\, (DFG/GSC 266). M.K. acknowledges the EU (INSPIN, FP7-ICT-X 612759; IFOX, NMP3-LA-2012 246102) and the Center of Innovative and Emerging Materials at Johannes Gutenberg University Mainz.

\bibliography{euler}
\bibliographystyle{apsrev}

\end{document}